\documentclass[sigconf, screen]{acmart}

\usepackage{framed}

\usepackage{booktabs} 
\usepackage{graphicx}
\usepackage{array}
\usepackage{url}
\usepackage{multirow}
\usepackage{color}
\usepackage{colortbl}
\usepackage{fancyhdr}
\usepackage{subfig}
\usepackage{diagbox}
\usepackage{placeins}
\usepackage{listings}
\usepackage{float}
\makeatletter
\def\UrlAlphabet{%
      \do\a\do\b\do\c\do\d\do\e\do\f\do\g\do\h\do\i\do\j%
      \do\k\do\l\do\m\do\n\do\o\do\p\do\q\do\r\do\s\do\t%
      \do\u\do\v\do\w\do\x\do\y\do\z\do\A\do\B\do\C\do\D%
      \do\E\do\F\do\G\do\H\do\I\do\J\do\K\do\L\do\M\do\N%
      \do\O\do\P\do\Q\do\R\do\S\do\T\do\U\do\V\do\W\do\X%
      \do\Y\do\Z}
\def\UrlDigits{\do\1\do\2\do\3\do\4\do\5\do\6\do\7\do\8\do\9\do\0}
\g@addto@macro{\UrlBreaks}{\UrlOrds}
\g@addto@macro{\UrlBreaks}{\UrlAlphabet}
\g@addto@macro{\UrlBreaks}{\UrlDigits}
\makeatother

\usepackage[linesnumbered, ruled,vlined]{algorithm2e}

\newcommand{\todo}[1]{}
\renewcommand{\todo}[1]{{\color{red} TODO: {#1}}}

\author{Shijian Chen}
\affiliation{
  \institution{Sun Yat-sen University}
  \city{Zhuhai}
  \country{China}
}
\email{chenshj73@mail2.sysu.edu.cn}

\author{Jiachi Chen}
\affiliation{
  \institution{Sun Yat-sen University \& The State Key Laboratory of Blockchain and Data Security, Zhejiang University}
  \city{Zhuhai}
  \country{China}}
\email{chenjch86@mail.sysu.edu.cn}
\authornote{Corresponding author.}

\author{Jiangshan Yu}
\affiliation{
  \institution{The University of Sydney}
  \city{Sydney}
  \country{Australia}
}
\email{jiangshan.yu@sydney.edu.au}

\author{Xiapu Luo}
\affiliation{
  \institution{The Hong Kong Polytechnic University}
  \city{Hong Kong}
  \country{China}}
\email{csxluo@comp.polyu.edu.hk}

\author{Yanlin Wang}
\affiliation{
  \institution{Sun Yat-sen University}
  \city{Zhuhai}
  \country{China}}
\email{wangylin36@mail.sysu.edu.cn}

\copyrightyear{2024}
\acmYear{2024}
\setcopyright{acmlicensed}\acmConference[Internetware 2024]{15th Asia-Pacific Symposium on Internetware}{July 24--26, 2024}{Macau, Macao}
\acmBooktitle{15th Asia-Pacific Symposium on Internetware (Internetware 2024), July 24--26, 2024, Macau, Macao}
\acmDOI{10.1145/3671016.3674808}
\acmISBN{979-8-4007-0705-6/24/07}

\begin{document}

    \title{The Dark Side of NFTs: A Large-Scale Empirical Study of Wash Trading}

    \begin{abstract}
NFTs (Non-Fungible Tokens) have seen significant growth since they first captured public attention in 2021. However, the NFT market is plagued by fake transactions and economic bubbles, e.g., NFT wash trading. Wash trading typically refers to a transaction involving the same person or two colluding individuals, and has become a major threat to the NFT ecosystem. Previous studies only detect NFT wash trading from the financial aspect, while the real-world wash trading cases are much more complicated (e.g., not aiming at inflating the market value). There is still a lack of multi-dimension analysis to better understand NFT wash trading. Therefore, we present the most comprehensive study of NFT wash trading, analyzing 8,717,031 transfer events and 3,830,141 sale events from 2,701,883 NFTs. We identify three types of NFT wash trading and propose identification algorithms. Our experimental results reveal 824 transfer events and 5,330 sale events (accounting for a total of \$8,857,070.41) and 370 address pairs related to NFT wash trading behaviors, causing a minimum loss of \$3,965,247.13. Furthermore, we provide insights from six aspects, i.e., marketplace design, profitability, NFT project design, payment token, user behavior, and NFT ecosystem.
\end{abstract}



    \maketitle
    \section{Introduction}
\label{Introduction}
NFTs (Non-Fungible Tokens)~\cite{wikinft} are blockchain-enabled digital assets, which users can buy and sell without third-party participation. Due to the rising enthusiasm for the concept, the NFT trading volume increased to an astonishing \$25.1 billion in 2021~\cite{online251B}. Many NFT marketplaces have been created to facilitate NFT trading. Take OpenSea, the NFT marketplace with over one million registered users~\cite{online1B}, as an example. It facilitated transactions (TXNs) for around \$5 billion alone in January 2022~\cite{online5B}. However, the market value of NFTs may not justify such a thriving market. According to Binance, almost 45\% of all NFT trading volume may be fraudulent due to wash trading events~\cite{online45}, where users manipulate the market by buying and selling the same financial product~\cite{onlinewhatiswashtrading}. A notorious case ~\cite{punk99980,punk99981,punk99982,punk99983,punk99984} involves the wash trading on \textit{Cryptopunk \#9998} (an NFT)\footnote{The reference of all the NFTs, TXN hashes, and Ethereum accounts mentioned in each section can be found at \href{https://github.com/NFTWashTrading/The_Dark_Side_of_NFTs/blob/main/REFERENCE.csv}{REFERENCE.csv}}, where the buyer initially obtained loans from multiple sources to purchase the NFT, then immediately sold the NFT to the origin holder for the same price, and finally repaid the loans. Another example is that \textit{Meebits \#13824} was traded twice between 0x35D0CA and 0xA99A76 for 14,700 WETH~\cite{WrappedEther}(a cryptocurrency) and 15,000 WETH, around 40 times the previous trading price.

The existence of NFT wash trading has been proven in the previous works~\cite{von2022nft,das2021understanding,serneels2023detecting,tariq2022suspicious,wen2023nftdisk}, with several features abstracted, e.g., von Wachter et al.~\cite{von2022nft} discovered that 2.04\% of NFTs' sale events trigger suspicions of market abuse, while Das et al.~\cite{das2021understanding} adopted the strongly/weakly connected component of the graph constructed by NFTs' events to detect NFT wash trading. However, they mainly abstracted NFT wash trading as graph patterns and only focused on its financial losses, while ignoring TXNs themselves (e.g., the trading price) and other aspects of the NFT ecosystem. Investigating NFT wash trading remains challenging due to the lack of a broader discussion on its definition, a researcher-friendly dataset, and more valuable insights. To fill the gap, we present the most comprehensive study on NFT wash trading, with findings from six aspects.

In this paper, we first collect NFTs' sale/transfer events through API access and related block/ERC-20 TXNs via open-source datasets. Then, we define three types of NFT wash trading, i.e., Round-trip Trading, Unprofitable Trading, and Hidden Trading. Next, we design heuristic algorithms to detect each type and adopt FP-Growth~\cite{han2004mining} to identify wash trading address pairs/groups. We flag 824 transfer events, 5,330 sale events, 370 address pairs, and 29 address groups related to wash trading, accounting for tokens worth around \$8,857,070.41. Based on the experimental results, we offer insights from six aspects, i.e., marketplace design, profitability, NFT project design, payment token, user behavior, and the NFT ecosystem. The main contributions of this paper are summarized as follows:
\begin{itemize}
    \item To the best of our knowledge, this work is the most comprehensive study on NFT wash trading. We identify three forms of NFT wash trading and propose heuristic algorithms for their detection.
    \item We contribute an extensive dataset of 2,701,883 NFTs' event sequences from 285 most popular collections. We have released the open-source processed datasets to help researchers uncover further studies\footnote{ \url{https://github.com/NFTWashTrading/The_Dark_Side_of_NFTs}}.  
    \item We systematically evaluate the NFT wash trading results, including their financial impact, trend, market liquidity, and insights from six aspects. In addition, we provide practical advice to NFT marketplaces.
\end{itemize}

    \section{Background}
In this section, we introduce the background knowledge of Ethereum and NFTs.
\label{background}
\subsection{Ethereum}

Blockchain is a decentralized, peer-to-peer network system that relies on cryptographic algorithms to secure data and consensus mechanisms to validate transactions. These technologies work together to ensure the integrity and transparency of the blockchain, making it a reliable system for various applications~\cite{wikiblockchain}. Based on blockchain technology, Ethereum is a distributed ledger platform with programmable features. With the design of smart contracts and the account-based model, Ethe\-reum allows people to initiate TXNs and develop applications for users to interact with on blockchains~\cite{wikiethereum}. It increases the diversity of the decentralized world, including the prosperity of NFTs.


\textbf{Smart contracts}~\cite{wikismartcontract} are tamper-proof programs stored on the blockchain that run when predetermined conditions are met. They facilitate TXNs in the decentralized system. Ethereum has two account types, i.e., \textbf{externally-owned account (EOA)} and \textbf{contract account}~\cite{official_account}. EOAs are controlled by anyone with private keys, while the contract account is associated with the smart contract code~\cite{vujivcic2018blockchain}. Accounts in Ethereum are anonymous but traceable. Moreover, a single user can hold multiple accounts without providing personal information.

\textbf{Ether (ETH)} is the native token circulating in Ethereum, used as a payment system for verifying TXNs~\cite{wikiethereum}. \textbf{ERC-20 token} is a standard for creating alternative tokens developed by smart contracts~\cite{official_erc20}.

A \textbf{block TXN}~\cite{Zheng2020} constructs the body of the block in Ethereum and refers to the TXN where the sender sends ETH or other tokens to the receiver with some additional information (e.g., smart contract function calls). An \textbf{internal TXN} refers to the TXN that occurs during the execution of a smart contract~\cite{chan2017ethereum}. When a smart contract is called, it may execute multiple functions, and each function may trigger calls to other contracts. The invocation and interaction between these contracts are achieved through internal TXNs. In this paper, we refer to all TXNs involving the transfer of ERC-20 tokens between addresses as \textbf{ERC-20 token TXNs}.

\begin{figure}[!htp]
  \begin{minipage}{1\linewidth}
    \centering
    \includegraphics[width=1\linewidth]{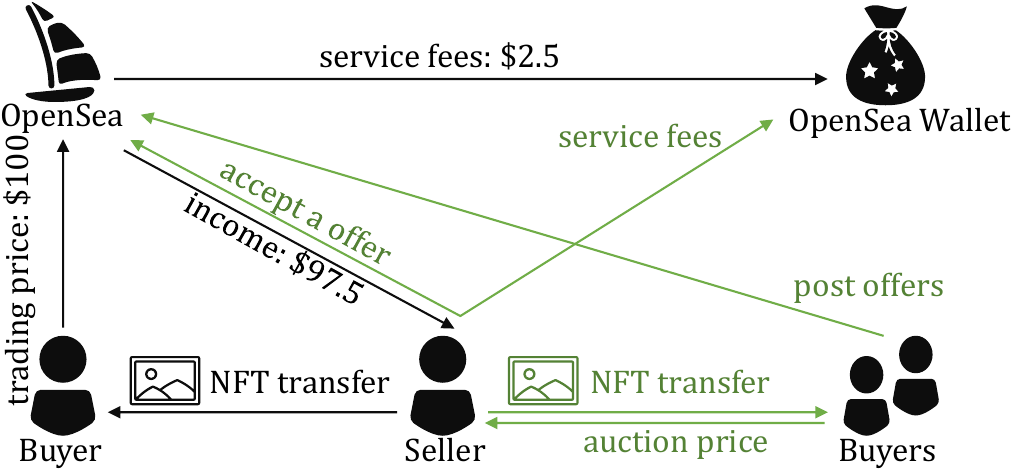}
    \caption{Two main ways to purchase NFTs on OpenSea: instant sale (Black), auction (Green).}
    \label{fig: sellandbuy}
  \end{minipage}
\end{figure}

\begin{figure}[!htp]
  \begin{minipage}{1\linewidth}
    \centering
    \includegraphics[width=1\linewidth]{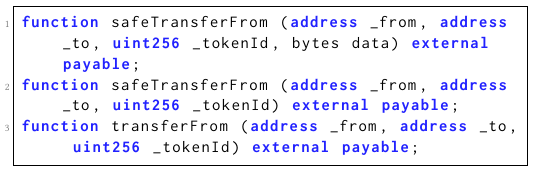}
    \caption{NFT transfer functions standardized by ERC721~\cite{official_erc721}}
    \label{fig: erccode}
  \end{minipage}
\end{figure}



\subsection{NFT Trading}

An NFT (Non-Fungible Token) is a unique on-chain identifier that typically refers to digital files, such as photos. Most NFTs are built using the ERC-721 standard~\cite{official_erc721}, which defines the primary interface for users to own, trade, and manage their tokens through a smart contract. NFTs are transferred on-chain using functions displayed by Figure \ref{fig: erccode}. The transfer of an NFT involves the \textit{from} and \textit{to} addresses, which represent the propagation of the NFT, and the \textit{tokenId} and contract address, which together represent a unique NFT on the blockchain\footnote{We only discuss ERC-721 NFT in this paper.}. Within a collection, each \textit{tokenId} is different. The first transfer of an NFT is also called the minted event, which shows the publishing of an NFT on the blockchain.


NFTs are often purchased on various trading platforms, such as OpenSea. Regarding OpenSea, as shown in Figure \ref{fig: sellandbuy}, there are two main ways to purchase an NFT. \textbf{1) }instant sale, where an NFT is listed at a fixed price, and a buyer can purchase it immediately, with a portion of the trading value being transferred to OpenSea's official wallet as service fees. \textbf{2) }auction, where buyers place bids on an NFT, and sellers can accept the offer. In this case, the sellers pay the service fees (part of the auction price) to OpenSea. The payment token for the auction is WETH~\cite{WrappedEther}.

    \section{WASH TRADING/TRADER TYPES}
\label{example}
In this section, we delve into the high-level ideas of three
wash trading types, providing detailed examples. Then, we define
wash-trading pairs/groups to flag wash traders and confirm
a related case of colluding addresses.


\subsection{Type 1: Round-trip Trading}
\textbf{Explanation. }Round-trip Trading is the unethical practice of repeatedly buying and selling the same securities to manipulate the finance market~\cite{onlinertt}. In the NFT market, Round-trip Trading happens when someone purchases an NFT and promptly resells it, either directly or through multiple addresses, to the original NFT holder. 

\noindent \textbf{Example. }\textit{OG:Crystal \#4015} was traded 24 times at almost the same price between 0xCF6FF6 and 0xC17D7c within 9 hours, accounting for 126.2 ETH. 

\subsection{Type 2: Unprofitable Trading}
\textbf{Explanation. }We define that Unprofitable Trading refers to an NFT TXN where the buyer is either funded beforehand by the seller or receives the seller's return amount, indicating that the TXN is not intended for profit. Normally, the transfer of NFT (seller $\rightarrow$ buyer) and the transfer of TXN amount (buyer $\rightarrow$ seller) occur simultaneously within a block TXN, which is guaranteed by smart contracts. However, Unprofitable Trading occurs with an additional value transfer, e.g., the seller transfers a certain amount of ETH that is similar to the NFT trading price to the buyer shortly before or after the TXN. Another conduction is using a certain amount of ERC-20 tokens with market value instead of ETH. An early value transfer can be interpreted as a form of funding, while the restitution of the amount illustrates that the NFT TXN does not profit from a third party. 

\noindent \textbf{Example. }For \textit{Omnimorph \#3980}, 0x6149ca (seller) transferred 0.23 ETH (104.5\% of the trading price) to 0x2beba3 (buyer) three minutes before the NFT sale event. Twenty minutes later, 0x2beba3 (new seller) returned 0.2\-098 ETH (95\% of the trading price) to 0x0f767ef
 (new buyer) four minutes after the next sale event. The two NFT sellers did not profit from either TXN, which was abnormal for a usual sale event. They conspired to complete the wash trading process. 
\subsection{Type 3: Hidden Trading} 
\textbf{Explanation. }Hidden Trading is third type of wash trading we identify, characterized by collusive actions between sellers and designated buyers. Hidden Trading implies the existence of a series of continuous private tradings where the NFT sellers designate the NFT buyers. On OpenSea, NFT holders can reserve items for specific buyers~\cite{onlineprivatesale}, i.e., only the addresses approved by the holders have the right to purchase the NFT. Private Tradings disregard the impact of market liquidity and user sentiment because the sellers and buyers know each other before the TXNs occur. 

\noindent \textbf{Example. }Wash traders raised the NFT price from 3.750 ETH, 3.780 ETH to 4 ETH, 7.4 ETH for \textit{VeeFriends \#7582}, all through continuous private tradings. 

\subsection{Wash trading pairs/groups}
\textbf{Explanation. }We define wash trading pairs as address pairs with high-frequency wash trading behaviors, while a wash trading group consists of multiple relevant wash trading pairs. Suspicious addresses will no longer be regarded as incidentally or deemed innocently involved in NFT wash trading if they frequently participate in Round-trip Trading, Unprofitable Trading, and Hidden Trading. Specifically, address pairs/groups involved in three types of NFT wash trading more than a certain number of times can be marked as wash trading pairs/groups. If there is a common address/TXN connecting both pairs, we assume that the two pairs come from the same wash trading group. 

\noindent \textbf{Example} \textit{Loot \#2157} was wash-traded ten times between by 0xB763\-9A and 0x996665 for 1 ETH between 2022-08-10T12\-:11:14Z and 2022-08-11T03:25:36\-Z. Shortly after, it was sold by 0xB7639A to 0xcc8990, and the same wash trading behavior happened again between 0xcc8990 and 0xBF1eD4, and more address pairs. Within a day, 178 transactions transpired back and forth at nearly the same price, and the domain names of sellers and buyers were highly similar, all in the form of Chinese license plates. Here, 0xB7639A and 0x996665, 0xcc8990 and 0xBF1eD can be treated as two wash trading pairs, while 0xB7639A, 0x996665, 0xcc8990, and 0xBF1eD is a wash trading group.

    \section{DATA COLLECTION}
\label{datacollectionpreprocessing}
In this section, we detail the methodology used to construct our dataset, which is derived from API access and open-source datasets.

\begin{table*}[!htp]
\footnotesize
\centering
\begin{tabular}{|c|c|c|c|c|c|c|c|c|c|}
\hline
\textbf{timestamp}&\textbf{collection}&\textbf{tokenId}&\textbf{from}&\textbf{to}&\textbf{type}&\textbf{isPrivate}&\textbf{payToken}&\textbf{numToken}&\textbf{usdToken}\\
\hline
2021-12-19T01:00:08&Alpha Shark&9&0x000000&0x1c2fd0&transfer&NaN&NaN&NaN&NaN\\
\hline
2021-12-21T15:59:40&Alpha Shark&9&0x1c2fd0&0x9164e3&transfer&NaN&NaN&NaN&NaN\\
\hline
2022-06-15T16:56:53&Alpha Shark&9&0x9164e3&0x99264d&sale&FALSE&ETH&2.9&1215.68\\
\hline
\end{tabular}
\caption{An example of event sequence: \textit{Alpha Shark \#9}'s event sequence.}
\label{Table: fields}
\end{table*}

\subsection{Event sequence}
The OpenSea API allows users to fetch metadata and core elements of NFTs~\cite{official_OpenSeaAPI}, e.g., sale/transfer events. The transfer events for each NFT reveal its circulation flow, while the sale events additionally provide specific trading information. We retrieve 8,717,031 transfer events and 3,830,141 sale events of 2,701,883 NFTs from Ethereum's 285 most popular collections. We start counting each NFT's event sequence since it was minted ($0$th record), and the $i$th record's fields will be marked as \textit{timestamp\_i}, etc. Table \ref{Table: fields} displays an example of the event sequence, i.e., since Alpha Shark (\textit{collection}) \#9 (\textit{tokenId}) was minted, it was transferred to 0x1c2fd0 (\textit{to\_0}), then to 0x9164e3 (\textit{to\_1}), then bought by 0x99264d (\textit{to\_2}) for 2.9 (\textit{numToken\_2}) ETH (\textit{payToken\_2}) at 2022-06-15T16:56:53 (\textit{timestamp\_2}), while each ETH cost \$1215.68 (\textit{usdToken\_2}). Also, \textit{isPrivate} indicates whether the NFT is reserved for a specific buyer.

\subsection{Block TXN and ERC-20 token TXN}
XBlock~\cite{xblock} is a data platform for the blockchain community. This data source~\cite{Zheng2020} provides researchers with information about block TXNs and ERC-20 token TXNs. To enable the identification of ETH and ERC-20 token transfers, we filter out 184,008,844 block TXNs and 48,513,194 ERC-20 token TXNs in the field of the same \textit{from} or \textit{to} as that of each NFT sale/transfer event.

\subsection{Historical market price data} 
CoinGecko API enables users to obtain crypto prices, historical market data, etc~\cite{official_coingecko}. We adopt it to collect the historical market price of ERC-20 tokens. Specifically, we identify 233,618 ERC-20 token smart contracts involved in related ERC-20 token TXNs. Since not all ERC-20 tokens have a market price reference value, we finally collect 2,373,787 historical price records of 2,982 ERC-20 tokens.

\subsection{Dataset overview}
Table \ref{Table:datacollection} demonstrates the final look of our dataset, and we make some preliminary calculations on it, e.g., holders of each NFT change on average 3.2 times each year, indicating that the NFT market does not enjoy high liquidity. Therefore, consecutive TXNs in a short period are abnormal and worth being concerned about. Also, the selected block TXNs and ERC-20 token TXNs contain ETH and ERC-20 token transfer information between each NFT sale event's \textit{from} and \textit{to}, exposing that the related addresses generate more behaviors before/after the NFT TXNs, as there is no direct transfer of ETH or ERC-20 token between two trading addresses for a common deal on OpenSea, except for the auction, using WETH as a payment token.

\begin{table}[!htp]
\centering
\resizebox{1\columnwidth}{!}{
\begin{tabular*}{\linewidth}{@{}l||l||c@{}}
\toprule
\small \textbf{Source} & \small \textbf{No. of} & \small \textbf{Statistic} \\
\midrule
\small OpenSea API& \small collection & \small285 \\
\small OpenSea API& \small NFT & \small2,701,883 \\
\small OpenSea API& \small address & \small902,571 \\
\small OpenSea API& \small sale event & \small3,830,141 \\
\small OpenSea API& \small transfer event & \small8,717,031 \\
\small XBlock& \small selected block TXN & \small184,008,844 \\
\small XBlock& \small selected ERC-20 token TXN & \small48,513,194 \\
\small CoinGecko API& \small ERC-20 token & \small2,982 \\
\small CoinGecko API& \small historical price record & \small2,373,787 \\
\bottomrule
\end{tabular*}
}
\caption{The final look of our dataset applied to the identification algorithm.}
\label{Table:datacollection}
\end{table}

\section{DETECTION}
\begin{figure*}[!hbt]
\centering
\includegraphics[width=0.75\textwidth]{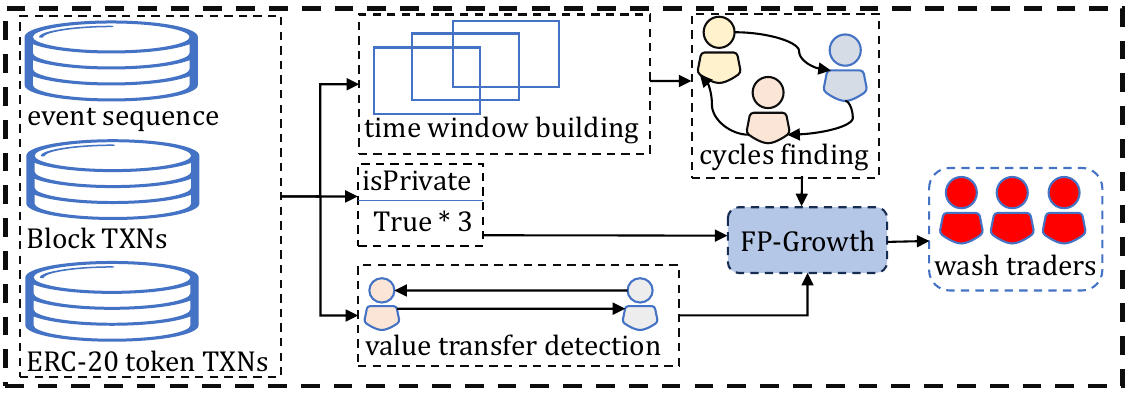}
\caption {The workflow of wash trading and wash trader identification}
\label{fig: algorithm}
\end{figure*}
\label{wash trading identification}
As demonstrated in Figure \ref{fig: algorithm}, by using NFTs' event sequences and selected block/ERC-20 token TXNs, we develop a series of heuristic algorithms to identify wash trading/traders. 
\subsection{Approach on Round-trip Trading}
\subsubsection{Approach design}
Our preliminary exploration of Round-trip Trading suggests that the NFTs are purchased repeatedly in a short time interval. Based on its definition, we build a multi-edge directed graph for each NFT's event sequence, where nodes represent addresses and edges denote the circulation of the NFT. Suspicious activities related to Round-trip Trading will be detected through cycles. During the identification period, a few challenges exist:
\begin{itemize}
\item \textit{Definition of a short time interval is ambiguous.} Since there is no empirical value for a time interval regarding how far apart two TXNs that are not suspected of wash trading should occur. To address this, we design a time window segment method.
\item \textit{Whether a transfer event contains trading information is uncertain.} It is possible that a transfer event is essentially a real NFT trading. To extend the detection, we have included transfer events in our algorithm.
\end{itemize}

\subsubsection{Time window} 
\label{subsec: time window}
We define a period of consecutive TXNs with a short time interval as a non-overlapping time window, and we set the following rules to segment the time windows. We use
\begin{equation}
    \textit{ATI}=\frac{\textit{timestamp}_{\textit{end}} - \textit{timestamp}_{\textit{start}}}{\textit{num}}
\end{equation}
as the threshold, where \textit{ATI} represents the average time interval between two adjacent events in each time window. For each time window, \textit{timestamp\textsubscript{end}} is \textit{timestamp} of the last record, while the \textit{timestamp\textsubscript{start}} is that of the first one, and \textit{num} denotes the number of the adjacent time intervals. If \textit{timestamp\_i} is \textit{timestamp\_\text{end}} of the current time window and $\textit{timestamp\_{(i+1)}} - \textit{timestamp\_i} \leq \textit{ATI}$, we consider the \textit{i+1}th record should be included in the time window; otherwise, the \textit{i+1}th\ record is the first record of a new time window. Although we disregard the relationships between TXNs separated for a long time, we care about the potential continuous Round-trip Trading. Thus, regardless of the \textit{ATI}, the \textit{i+1}th record will be included if \textit{to} of the \textit{i+1}th record has ever occurred in the current time window. We initialize the \textit{ATI} with an empirical value of 84,400 seconds (one day)\footnote{Choosing different values for the key parameters in our algorithm produces different results. We actually provide a value reference while guaranteeing the correctness of the results and allowing researchers to adjust them according to datasets of various sizes. See how we determine the value from \href{https://github.com/NFTWashTrading/The_Dark_Side_of_NFTs/blob/main/thresholdExplanation.md}{here}.\label{plz}}.

\subsubsection{Cycles finding}
For each collection, we represent the graph constructed by event sequences in each time window of an NFT with a unique identifier, \textit{G\textsubscript{ij}}, where \textit{i} is the NFT's \textit{tokenId} and \textit{j} is the index of the time window that starts from 0. In \textit{G\textsubscript{ij}}, $\textit{v}\in \textit{V}$ is the set of addresses, and $\textit{e}\in \textit{E}$ indicates the NFT's propagating direction. We record all edges going in the same direction into a dictionary \textit{D}, where the key is a tuple of the start and end nodes of each edge, and the value is the list of edges with the same direction. We adopt Depth First Search (DFS)~\cite{tarjan1972depth} to find all cycles for \textit{G\textsubscript{ij}} and establish rules to confirm if Round-trip Trading exists by investigating the potential paths from \textit{D}.

\subsubsection{Wash trading confirmation}
If a cycle's \textit{D} performs in \textit{{(v\textsubscript{1}, v\textsubscript{2}): [e\textsubscript{1}, e\textsubscript{2}]; (v\textsubscript{2}, v\textsubscript{3}): [e\textsubscript{3}]; (v\textsubscript{3}, v\textsubscript{1}): [e\textsubscript{5}, e\textsubscript{6}]}}, the total number of suspicious walks for the cycle is 2 * 1 * 2 = 4. To prevent misjudgment and ensure stringent identification, each cycle is confirmed as Round-trip Trading behavior only when one of the following rules is met. \textbf{1) }We consider a cycle included in Round-trip Trading when its number of repetitive walks is no less than the strict threshold, 10 * 10 = 100\footref{plz}, meaning at least ten back-and-forth tradings happen for two addresses. \textbf{or 2) }If all the events in one of the walks are sale events, the cycle would be identified as Round-trip Trading. 

\subsection{Approach on Unprofitable Trading}
ETH and ERC-20 token transfers constitute additional significant evidence for distinguishing wash trading behavior. We establish three rules to detect each Unprofitable Trading behavior. \textbf{1)} \textit{from} and \textit{to} of the related ERC-20 token/block TXNs are the same as the corresponding sale event. \textbf{2)} Block TXNs with null input data are considered, meaning that they only contain ETH transfers without any invocation of smart contracts. Specifically, the time threshold is 20 minutes\footref{plz}, i.e., only ETH transfers that occurred 20 minutes before/after the sale event will be included. \textbf{3)} For ERC-20 token transfer, the time threshold is 80 minutes\footnotemark[3], i.e., only ERC-20 token transfers that occurred 80 minutes before and after the sale event will be included.
\subsection{Approach on Hidden Trading}
We use \textit{isPrivate} of each record to indicate whether trading is private. For each NFT's event sequence, each group of three and more continuous private trading would be detected as Hidden Trading behavior. 
\subsection{Detection of wash trader}
The probability of addresses appearing simultaneously as a pair or a group provides the reference for wash trader detection. We utilize FP-Growth~~\cite{han2004mining}, an association-rule mining algorithm, on addresses related to Round-trip Trading, Unprofitable Trading, and Hidden Trading, to find the suspicious address pairs. We define the tuple \textit{R: (from, to)} for Round-trip Trading, \textit{U: (from, to)} for Unprofitable Trading, to represent the wash trading pairs. Considering the colluding relationship is not displayed by each sale event of Hidden Trading, we include every address of the continuous private trading, i.e., the tuple \textit{H: (address1, address2, address3, ...)}. Given a set of addresses, frequent pattern mining can find all the itemsets with a frequency greater than the support. The matrix \textit{$[R_{all}; U_{all}; H_{all}]^T$} is the final input for FP-Growth, where the elements for each row vector come from the tuples of \textit{R}, \textit{U}, or \textit{H}. If the address pair/group's occurrence frequency exceeds the support, the address pair/group is detected as a wash trading pair/group. Also, we combine all wash trading pairs with at least one common address/TXN  between each other as a wash trading group.

    \section{EXPERIMENTAL RESULTS}
\label{experimental result}
This section demonstrates the results of our identification methodology for NFT wash trading and wash traders.
\begin{table}[H]
  \begin{center}
  \resizebox{\linewidth}{!}{
    \begin{tabular}{l|c|r}
    \hline
    \textbf{Collection} & \textbf{No. of behaviors} & \textbf{Rank of Market value}\\
      \hline 
      \textit{OG:Crystal} & 9412 & 204th\\
      \textit{Apes R Us} & 99 & 237th\\
      \textit{Meebits} & 94& 9th\\
      \textit{Bored Ape Yacht Club} & 57 & 2th\\
      \textit{hashmasks} & 45 &  26th\\\hline      
    \end{tabular}
    }
    \caption{Top five collections by the number of wash trading behaviors}
    \label{table: top5}
  \end{center}
\end{table}

\subsection{Results}
\subsubsection{Results overview}We identify 824 transfer events and 5,330 sale events related to NFT wash trading, accounting for \$8,857,070.41, which is around 0.12\% of the total trading amount (\$7,216,023,387.61). It suggests that the financial impact of NFT wash trading on popular collections is relatively small and overrated. However, as shown in Table \ref{table: top5} that displays the top five collections with their market value ranked by the number of wash trading behaviors, \textit{OG:Crystal} with lagging ranking is wash traded most, far exceeding the other collections. It indicates the harm of wash trading inflicted on the NFT market might not be fully exposed through trading volume, inspiring us to explore broader aspects of its impact.

\subsubsection{Results of Round-trip Trading}\label{subsubsec: rtt_results}Our results show that 3,041 Round-trip Trading events from 2,948 time windows are detected, accounting for \$6,453,831.51. We find each NFT has around one wash-traded time window, and it supports our time window division, i.e., continuous Round-trip Trading behaviors are detected together. For instance, 0x837E6f and 0xABE3aE wash traded \textit{Bean \#16738} in multiple time periods, but eventually, the events involved are included in a single time window, containing 14 TXNs. 

\subsubsection{Results of Unprofitable Trading}The results include two parts. \textbf{1) }We identify 2,238 sale events with suspicious ETH transfers, accounting for \$2,247,91\-5.26. 59.30\% of sale events occur when the NFT sellers transfer ETH to the NFT buyers to fund the purchase activities 20 minutes before the trading, while the remaining events happen as the NFT sellers return ETH to the NFT buyers 20 minutes after. \textbf{2) }There are 768 sale events with suspicious ERC-20 token transfers, accounting for \$1,709,390.72. Among them, the most used ERC-20 token for Unprofitable Trading is WETH, with 99.3\%.

\subsubsection{Results of Hidden Trading}We detect 968 groups with three or more continuous private trading, including 4,257 sale events and accounting for \$1,556,267.23. Besides, NFTs' prices rise in 62.29\% of Hidden Trading cases, where 70.81\% of them even maintain an upward trend in all private trading, e.g, the price of \textit{Bored Ape Yacht Club \#5332} goes up from 3.55, 5.63, 6.5432 to 8.95 ETH all through Hidden Trading. However, as we search for segments of three or more adjacent sale events with continuously increasing prices in each NFT's event sequence, we find less than 5\% of the NFTs meet the requirement. It suggests wash traders may adopt Hidden Trading to inflate the prices of NFTs.

\begin{figure}[!htp]
\begin{minipage}{1\linewidth}
\centering
\includegraphics[width=1\linewidth]{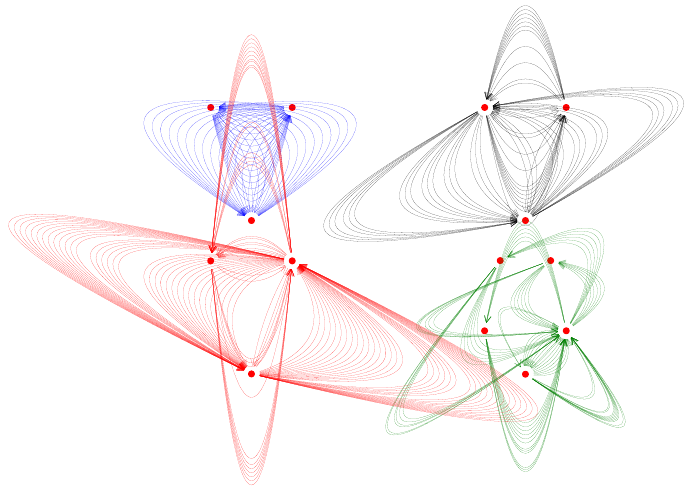}
\caption{Obvious evidence in visualization for wash trading groups.}
\label{fig: addr}
\end{minipage}
\end{figure}

\subsubsection{Results of wash trading pairs/groups}We identify 370 wash trading pairs and 29 wash trading groups, indicating these addresses occur at least 24,311 * 0.00\-05 = 12 times, where 24,311 is the number of row vectors in our FP-Growth matrix, and 0.0005 is the support. To observe whether our results present obvious evidence of suspicious activities for users, we visualize all the related events of addresses from each wash trading group. As shown in Figure \ref{fig: addr} (the partial result of the visualization), each node marked in red represents an address from a group, and each directed edge between nodes represents the propagation of the NFT. It demonstrates obvious conspiracy in the wash trading groups.

\subsubsection{Trend of wash trading events.}
\begin{figure}[!htp]
\begin{minipage}{1\linewidth}
\centering
\includegraphics[width=1\linewidth]{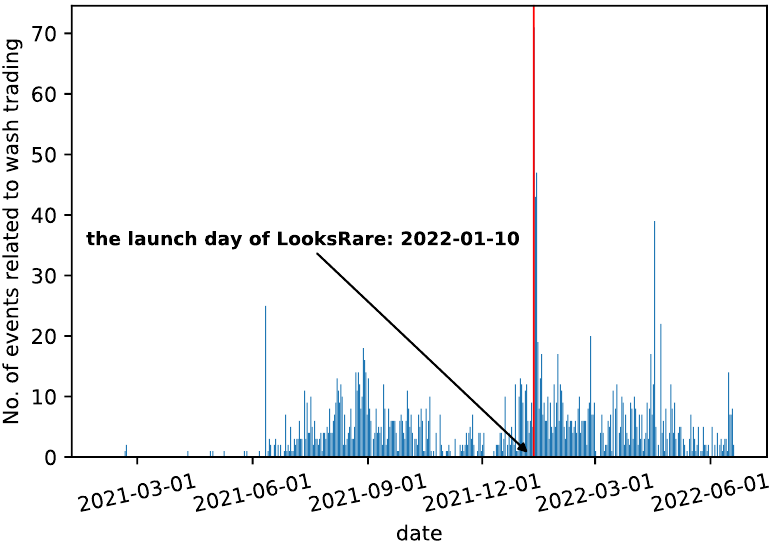}
\caption{The trend for the number of events related to wash trading, excluding \textit{OG:Crystal}.}
\label{fig: looksrare}
\end{minipage}
\end{figure}

We investigate the trend for sale/transfer events related to NFT wash trading. Particularly, we exclude \textit{OG:Crystal} to avoid its extremely leading role in our trend assessment. Figure \ref{fig: looksrare} shows the change in the number of wash trading events of our results from 2021-06-28T15:00:34Z to 2022-06-20:18:28:10Z. The events have grown and become stable since June 2021 and reach their peak on January 11, 2022. It indicates that wash trading has been used consciously after June 2021. For the peak, it is mostly because of the launch of \textit{LooksRare}~\cite{official_loosrare}, an NFT marketplace with an incentive reward plan that encourages frequent TXNs~\cite{onlinelooksrare} and thus spawns a bunch of wash trading behaviors~\cite{serneels2023detecting, la2022nft, cho2023non} (See \ref{finding: looksrare} for details).

\begin{figure}[!htp]
\begin{minipage}{1\linewidth}
\centering
\includegraphics[width=1\linewidth]{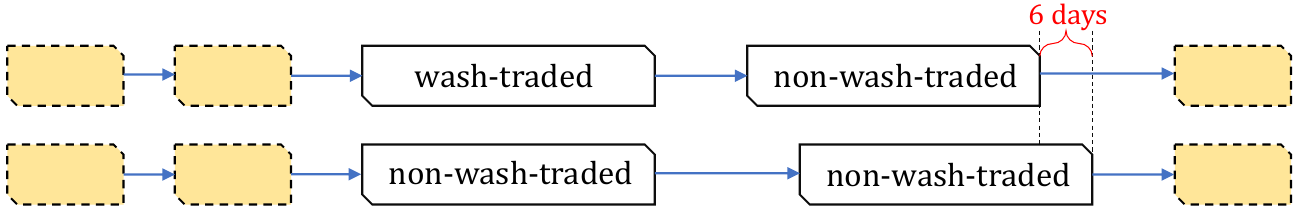}
\caption{Time interval differences when transitioning to the next non-wash-traded time window between wash-traded and non-wash-traded time windows}
\label{fig: timewindow}
\end{minipage}
\end{figure}

\subsubsection{Market liquidity.}
We then study the impact of NFT wash trading on market liquidity from the perspective of time windows. The interval of adjacent time windows provides evidence that Round-trip Trading stimulates NFTs sale events. It takes an average of 20.80 days for the current wash-traded time window to transition to the non-wash-traded one. In contrast, as shown in Figure \ref{fig: timewindow}, it takes six days longer in 1,725,783 cases of the current non-wash-traded time window, potentially accelerating NFT liquidity.

    \section{FINDINGS FROM MULTIPLE DIMENSIONS}
\label{sec: findings from multiple dimensions}
In this section, we provide an in-depth analysis of six key aspects related to NFT wash trading, including marketplace design, profitability, NFT project design, payment token, user behavior, and the NFT ecosystem.
\subsection{Marketplace design}
\label{finding: looksrare}
As shown in Figure \ref{fig: looksrare}, the NFT wash trading trend of events peaks on January 11, 2022. Noticeably, \textit{LooksRare} was released one day before. We investigate the collection (named \textit{Meebits}) with the most wash trading events on the peak and find that nearly 93\% of its TXNs occur on \textit{LooksRare}. It suggests that the wash trading may lead by the launch of \textit{LooksRare}. One of \textit{LooksRare}'s policies is rewarding the users who trade on the platform with \textit{LOOKS} (its native token). For example, \textit{Bored Ape Yacht Club \#4937} was sold at 91 ETH, and both the seller and buyer received around \$3.5K worth of \textit{LOOKS}. According to our experiment results, the NFTs from \textit{Meebits} associated with the wash trading peak were all sold for over 20 ETH, driven by the marketplace incentive policies. 

\begin{framed}
    \noindent\textit{\textbf{Finding 1:} Policies implemented by NFT marketplaces can encourage wash trading behaviors.}
\end{framed}




\subsection{Profitability}
Existing research has not yet explored whether wash traders can profit from benign users or measured the potential loss of the victims under the influence of wash trading. To fill the gap, we conduct the following analysis.

\begin{figure}[!htp]
  \centering
  \begin{minipage}[t]{0.48\linewidth}
    \centering
    \includegraphics[width=\linewidth]{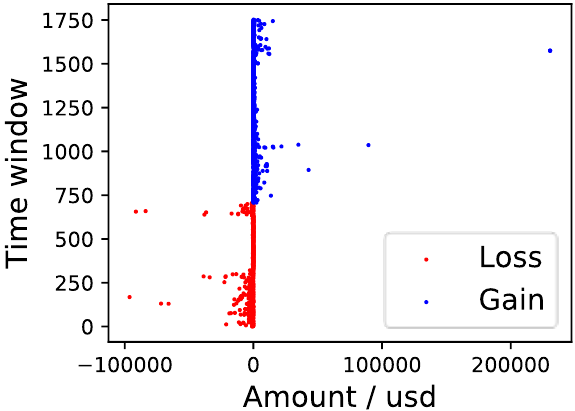}
    \caption{The distribution of all wash-traded time windows in terms of gain or loss}
    \label{fig: profit1}
  \end{minipage}
  \hfill
  \begin{minipage}[t]{0.48\linewidth}
    \centering
    \includegraphics[width=\linewidth]{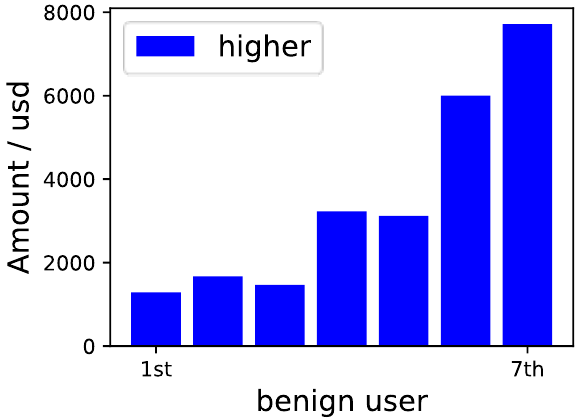}
    \caption{The difference between the last trading price in wash-traded time window with each trading price for the following 1st to 7th users}
    \label{fig: profit2}
  \end{minipage}
\end{figure}

We investigate the possibility for wash traders to profit from benign users without considering incentive rewards from NFT marketplaces through the results of Round-trip Trading. In other words, we assess if wash traders can offset the service fees (charged by NFT marketplaces for their service) by reselling the NFTs. Our analysis focuses on a scenario where a non-wash-traded time window follows a wash-traded time window. At the same time, the total service fees from round-trip sale events in the previous time window are less than or equal to the first sale event in the latter time window. To strengthen our analysis, we adopt OpenSea's service fee rate (2.5\%), the highest among major NFT marketplaces. Our analysis identifies 1,752 wash-traded time windows that satisfy the scenario. Among them, wash traders in around 60\% of cases achieve profitability when jumping to the next non-wash-traded time window. Figure \ref{fig: profit1} shows the distribution of all wash-traded time windows regarding gain or loss, where most are concentrated around 0, indicating that most wash traders do not suffer significant losses after deducting the service fees. For example, in the case of \textit{OG:Crystal \#895}, two traders (0x89a09c and 0x0B7742) traded the NFT six times with prices ranging from 0.06 ETH to 0.09 ETH during the first time window. The NFT was finally sold for 1 ETH in the first sale event of the second time window, resulting in a profit of 0.99 ETH for wash traders.



In addition, to explore the resale prices of NFTs after being wash traded, we compare the price of the final sale event in a wash-traded time window with that of the first one in the next non-wash-traded time window. The result demonstrates that 48.62\% of wash traders resell the NFT at a lower price when transitioning into the next time window, whereas this percentage is 36.86\% for ordinary users. This implies that wash traders tend to list NFTs at a lower price to accelerate sales. Moreover, to measure the impact on benign users in non-wash-traded time windows after each wash-traded time window, we compare the following trading price for users with the last trading price in wash-traded time window. For all 1,752 cases, there are up to seven TXNs after the wash-traded time window. Figure \ref{fig: profit2} shows the average difference between the trading prices. The following users sell their NFTs at a higher price as the TXNs increase, indicating that wash trading behaviors help to raise the overall price trend. If we treat the difference as the loss of each user, this portion of Round-trip Trading causes at least a total loss of \$3,965,247.13 for users.

\begin{framed}
    \noindent\textit{\textbf{Finding 2:} Most wash traders do not suffer significant losses or even profit from reselling NFTs to benign users after conducting Round-trip Trading. They tend to list NFTs at lower prices to attract more victims and accelerate sales.}
\end{framed}




\subsection{NFT project design}
\begin{figure}[!htp]
\begin{minipage}{1\linewidth}
\centering
\includegraphics[width=1\linewidth]{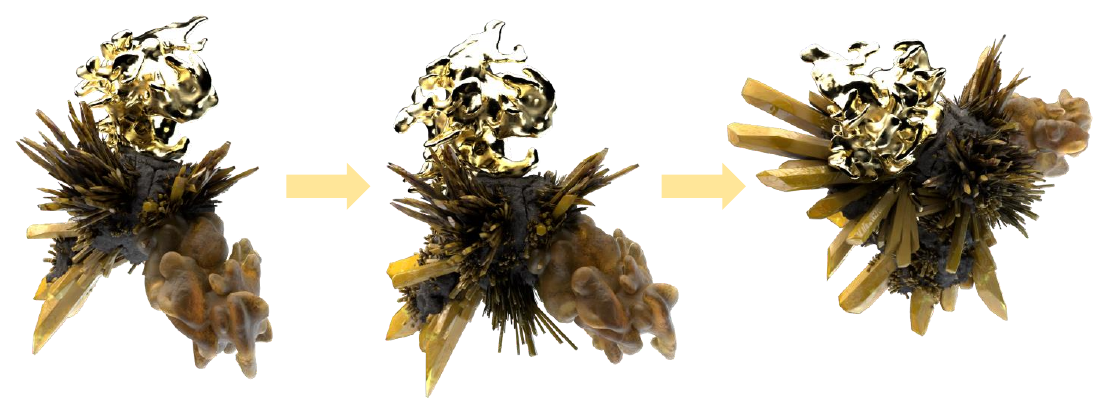}
\caption{The evolution of \textit{OG:Crystal \#1861} from the 5th generation to its final form through Round-trip Trading~\cite{og1861}}
\label{fig: og}
\end{minipage}
\end{figure}

As Table \ref{table: top5} shown, \textit{OG:Crystal} is the collection with the most wash trading behaviors totaling 9,412 times, accounting for \$1,031,4\-42.16. However, the main purpose of wash trading on \textit{OG:Crystal} is not to inflate trading volume but to enhance the appearance of the NFTs. The design and working mechanism of \textit{OG:Crystal} is responsible for this~\cite{ogwork,lagame}: each \textit{OG:Crystal} transforms and evolves with each purchase made by a new collector. The appearance of each piece is generated based on the properties of the owner's crypto wallet. Specifically, each \textit{OG:Crystal} will be `locked', meaning that new TXNs will not affect its shape, structure, or rarity, two months after the initial sale date or after it has grown by seven generations through multiple trading. For example, Figure \ref{fig: og} shows the evolution of \textit{OG:Crystal \#1861} from 5th generation to 7th generation through Round-trip Trading between 0x9FFFB\-8 and 0xB05Cb1. The overview result on \textit{OG:Crystal} is more convincing for verifying its working mechanism. \textbf{1)} Wash traders of \textit{OG:Crystal} generally appear as a pair, and the average number of walks formed by TXNs of each NFT is nearly 35, meaning that they may resell the NFT to each other around six times. \textbf{2)} Wash trading does not exist after the NFT is `locked' in 95.73\% cases. Both findings suggest that the design of \textit{OG:Crystal} can lead to wash trading problem, turning the project's vision that creates a public work of art for all users into personal wash trading activities. 

On the contrary, \textit{Kaiju Kingz} is a collection that exhibits minimal wash trading behavior, with a deliberate feature of its design~\cite{KaijuKingz}. Owners of \textit{Kaiju Kingz}'s NFTs can generate 5 \textit{RWASTE} per day, the circulating token for \textit{Kaiju Kingz} used for breeding. They can create a so-called \textit{Kaiju Kingz} baby (a new NFT) with 750 \textit{RWASTE}. In this sense, owners will be more inclined to hold the NFTs for a longer period of time.

Moreover, we manually investigate the design mechanisms of all collections with wash trading, but find no more factors that drive TXNs and potentially facilitate wash trading. It illustrates that \textit{OG:Crystal} is a unique presence in the NFT market and deserves the attention of subsequent NFT designers.

\begin{framed}
    \noindent\textit{\textbf{Finding 3:} The operational mechanism of an NFT project can be designed to encourage wash trading activities.}
\end{framed}


\subsection{Payment token}

\begin{table}[H]
\setlength{\fboxsep}{0pt}
\resizebox{\linewidth}{!}{
\setlength{\tabcolsep}{1pt}
\begin{tabular}{@{}l|l|l@{}}
\toprule
\textbf{time (+UTC)} & \textbf{event} & \textbf{value}\\ \midrule
\cellcolor[HTML]{EFEFEF}Aug-26-2021 05:46 AM & \cellcolor[HTML]{EFEFEF}WETH transfer: 0xcF5e38 $\rightarrow$ 0xe67753 & \cellcolor[HTML]{EFEFEF}0.1\\
\cellcolor[HTML]{EFEFEF}Aug-26-2021 05:49 AM & \cellcolor[HTML]{EFEFEF}WETH transfer: 0xcF5e38 $\rightarrow$ 0xe67753 & \cellcolor[HTML]{EFEFEF}0.07\\
\cellcolor[HTML]{EFEFEF}Aug-26-2021 05:50 AM & \cellcolor[HTML]{EFEFEF}WETH transfer: 0xcF5e38 $\rightarrow$ 0xe67753 & \cellcolor[HTML]{EFEFEF}0.3\\
\cellcolor[HTML]{EFEFEF}Aug-26-2021 05:52 AM & \cellcolor[HTML]{EFEFEF}WETH transfer: 0xcF5e38 $\rightarrow$ 0xe67753 & \cellcolor[HTML]{EFEFEF}0.47\\
\cellcolor[HTML]{C0C0C0}Aug-26-2021 05:56 AM & \cellcolor[HTML]{C0C0C0}Offer made: 0xe67753                           & \cellcolor[HTML]{C0C0C0}0.1\\
\cellcolor[HTML]{C0C0C0}Aug-26-2021 05:58 AM & \cellcolor[HTML]{C0C0C0}Offer accepted: 0xcF5e38                       & \cellcolor[HTML]{C0C0C0}0.1\\ 
\bottomrule
\end{tabular}
}
\caption{An example of Unprofitable Trading using ERC-20 token: \textit{Chibi Dino \#5723}}
\label{table: erc20}
\end{table}

We haven't discovered various ERC-20 tokens involved in Unprofitable Trading based on the results, and the main circulating ERC-20 token involved is WETH, accounting for 99.3\%. WETH enables users to submit pre-authorized bids that will be automatically fulfilled later without the bidder's permission. In our results, WETH mainly appears in two places. \textbf{1)} NFT holders invoke the smart contract, \textit{Wrapped Ether}, to transfer WETH to NFT bidders, and afterward, NFT holders accept NFT bidders' offer to obtain WETH back through the auction. For example, Table \ref{table: erc20} demonstrates the period of how 0xcF5e38 and 0xe67753 execute Unprofitable Trading: before the bidder 0xe67753 made an offer of 0.1 WETH to buy \textit{Chibi Dino \#5723}, 0xcF5e38 transferred total 0.1 + 0.07 + 0.3 + 0.47 = 0.94 WETH to 0xe67753. Two minutes after, 0xcF5e38 accepted the offer. It is a complete self-directed purchase that contributes to false trading volume. \textbf{2)} Another occurrence shows in the situation that there is another NFT sale event with the opposite \textit{from} and \textit{to} as the current one 80 minutes before. The WETH was transferred from \textit{from} to \textit{to} through the previous bidding auction and is transferred back through the current one. It is Round-trip Trading with multiple bids won, e.g., 0xeB1543 and 0xB893AE accept each other's offer respectively on \textit{OG:Crystal \#5609} and \textit{OG:Crystal \#6255} within one minute, and their offer prices are the same, i.e., 0.13 WETH.

During the auction process, we notice more obvious evidence of address collusion regarding WETH, e.g., some auction is essential almost free transfer of NFT, making the NFT's historical price curve sag seriously and creating a huge deviation from the market reference value. For example, the accepted offer price for \textit{dotdotdot \#2663} on December 30, 2021, is less than 0.0001 WETH, while its previous trading price is 3 ETH. The NFT holder and bidder know each other for such abnormal behavior, and their user names all start with \textit{blitmonk}. To explore the observation, we define $\textit{PF}=\frac{\textit{Price}{\textit{current}}}{\textit{Price}{\textit{latest}}}$ to represent the trading price fluctuations, where we start counting from 0 for each NFT's all sale events, i.e., the \textit{0}th\textit{PF} is the price fluctuation between an NFT's first and second sale events. To demonstrate server price sag, we specify that $\frac{1}{i\mathrm{th}\textit{PF}}$ and \textit{(i+1)}th\textit{PF} should be greater than 1,000. The results show 231 cases, only accounting for \$366.12. It suggests that this suspicious auction, which is overlooked by previous works, has been conducted on a small scale, and its collusive feature for the addresses involved should be taken into research.

\begin{framed}
    \noindent\textit{\textbf{Finding 4:} WETH needs to be focused most on applying payment tokens on NFT wash trading.}
\end{framed}




\subsection{User behavior}

\begin{figure}[!htp]
\begin{minipage}{1\linewidth}
\centering
\includegraphics[width=1\linewidth]{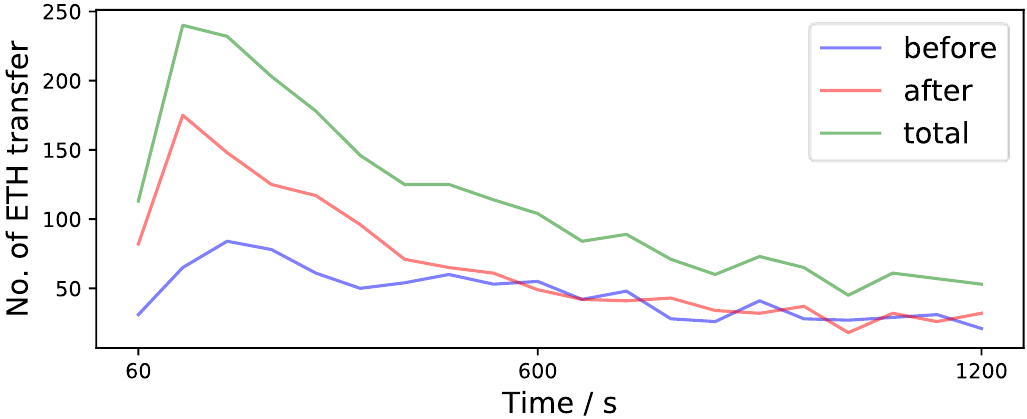}
\caption{The total number of ETH transfer events at different times before and after an NFT sale event.}
\label{fig: eth}
\end{minipage}
\end{figure}

In Unprofitable Trading, sellers collude with buyers to create unprofitable TXNs. For example, just 4 seconds before the sale event 0x1a4521, 0x4EDE98 completed a transfer (0.96 ETH) fully sufficient to support 0xE1f14d in buying \textit{Animeta \#4302} (0.435 ETH) in a normal TXN 0x53\-6fa8. To understand to what extent ETH transfer helps with each sale event, we define $M_{r}$ as the total amount received by sellers, $M_{t}$ as the total amount returned by sellers, and $M_{s}$ as the related NFT trading price. We then adopt Pearson correlation coefficient~\cite{cohen2009pearson} to measure the strength of the linear relationship between ($M_{r}$, $M_{s}$) and ($M_{t}$, $M_{s}$). The results are 0.7999 for ($M_{r}$, $M_{s}$) and 0.9840 for ($M_{t}$, $M_{s}$), indicating that the amount of each ETH transfer is similar to the NFT trading price. Furthermore, we count the number of ETH transfer events detected under a finer granularity (i.e., 60 seconds). Figure \ref{fig: eth} shows that the closer the time point before/after a sale event is, the more ETH transfer behaviors happen. Also, wash traders make the most frequent ETH transfer between two and four minutes before/after a sale event, after which the number gradually decreases. The respective high price and time correlations prove that these ETH transfers are directly for the NFT TXNs. 

\begin{framed}
    \noindent\textit{\textbf{Finding 5:} Users can adopt ETH transfers with the intention of directly servicing wash trading.}
\end{framed}





\subsection{NFT ecosystem}
\begin{table}[!htp]
  \begin{center}
  \resizebox{\linewidth}{!}{
    \begin{tabular}{l|c|c|r} 
    \hline
    \textbf{form} & \textbf{amount / before} & \textbf{amount / after} & \textbf{decrease}\\
      \hline 
      Round-trip Trading & \$6,453,831.51 & \$ 4,298,251.79 & 33.40\%\\
      Unprofitable Trading & \$2,247,915.26 & \$606,487.54 & 73.02\% \\
       Hidden Trading & \$1,556,267.23 & \$1,100,436.56 & 29.29\% \\
      \hline
    \end{tabular}
    }
    \caption{Statistical results of decreased wash trading behavior}
    \label{table: decrease}
  \end{center}
\end{table}
To better understand the impact of wash trading pairs/groups on NFT wash trading, we identify all 10,594 TXNs related to these address pairs and exclude their trading volume. For instance, if there is a flagged TXN in Hidden Trading, we do not calculate the trading volume of all consecutive private trading. As shown in Table \ref{table: decrease}, the results indicate a decrease of around 30\% in wash trading amount for round-trip and Hidden Trading, while Unprofitable Trading experiences a significant decline, indicating the prevalence of wash traders in Unprofitable Trading. Interestingly, the addresses involved in wash trading pairs/groups accounted for only 0.082\% of the total, meaning that a small group of users could cause considerable wash trading, and sizable wash trading behaviors could be eliminated by identifying and flagging them. It suggests that to avoid wash trading better and alert the users, NFT marketplaces do not need to keep all the trading addresses under governance but mark a small proportion of malicious accounts in event sequences.

\begin{framed}
    \noindent\textit{\textbf{Finding 6:} A small proportion of users could cause considerable wash trading amount.}
\end{framed}




    \section{DISCUSSION}
\label{lab:discussion}
\subsection{Threats to validity}
\subsubsection{Internal validity}
\textbf{1) }Adjusting the support for FP-Growth may affect the experimental results, e.g., the number of wash trading pairs/groups would increase if it is set loosely. Moreover, the support should be reconsidered for different datasets. To ensure the reliability of our results, we set a strict value, i.e., address pairs will be considered suspicious only when they participate in wash trading at least twelve times. \textbf{2) }Our method may ignore unknown wash trading. We have not yet investigated more complex trading networks, e.g., incorporating internal TXNs to understand the intention of NFT traders from a more underlying perspective and thus reveal more NFT wash trading patterns. However, we are still the most comprehensive study on NFT wash trading.
\subsubsection{External validity}
The current tools (e.g., OpenSea API) for collecting NFT events have shortcomings, such as mislabeling transfer/sale/minted events. Nevertheless, we address the related issues and make public a guaranteed dataset for researchers.\footnote{We have open-sourced our data, both before and after processing in \url{ https://drive.google.com/drive/folders/1bddfHZgk3BSmDUN0aTAub7mJ-_1_36ff}, which includes transfer and sale events for 285 NFTs collections.}

\subsection{Implications for NFT Marketplace}
 We find no restriction or warning in place to limit wash trading on major NFT marketplaces. For instance, OpenSea only reports NFTs involved in suspicious activities and compromised accounts but does not provide explanations for these reports. Alarmingly, we manually investigate the NFTs involved in wash trading but find no alerting information. Interestingly, OpenSea has altered the way it records private trading in NFT event sequences, switching from the generic \textit{`Sale'} to \textit{`Sale - Reserved'}. This positive change offers NFT participants more insight into the nature of the trade. Our algorithm has the potential to detect wash trading activities promptly by analyzing the historical trail of an NFT. It would benefit the overall health of the NFT ecosystem if our technology could be integrated into the front end of major NFT marketplaces to send alerts about wash trading when players purchase NFTs.

    \section{RELATED WORK}
\label{related work}
\subsection{Data collection for NFT research.}
NFT researchers mainly collect data in an external way. Von Wachter et al.~\cite{von2022nft} utilized OpenSea API to collect NFTs' events and adopted Coingecko API to retrieve the historic USD prices for crypto. White et al.~\cite{white2022characterizing} utilized a moving window of Unix timestamps to retrieve 5,252,252 sale events via OpenSea API. Apart from the API access, Das et al.~\cite{das2021understanding} collected more data through web scraping. 
\subsection{Detection for NFT wash trading.}Several studies have been published to detect wash trading behavior for crypto assets. In terms of Bitcoin, Aloosh et al.~\cite{aloosh2019direct} provided direct evidence for `fake volume' that occurred in cryptocurrency exchanges through trading records lea\-ked by hackers. For NFTs, current researchers pay more attention to NFT TXN networks rather than discussing different types of wash trading, including evidence from block/ERC-20 TXNs and Private Trading. For example, von Wachter et al.~\cite{von2022nft} presented methods to identify the graph patterns for wash trading, indicating that 2.04\% of sale events are suspicious. Based on that, La et al.~\cite{lagame} explored more suspicious graph patterns. Das et al.~\cite{das2021understanding} focused on SCC and WCC to discover the wash trading behavior. Wen et al.~\cite{wen2023nftdisk} provided a novel visualization method for identification. In addition to La et al.~\cite{lagame} who analyzed the profitability of wash trading, we are the first to place a broader discussion on forms/impacts of NFT wash trading, instead.

    \section{Conclusion and Future Works}
This paper introduces currently the most comprehensive understanding of NFT wash trading. We first organize a dataset containing 8,717,031 transfer events and 3,830,141 sale events from 2,701,883 NFTs of the 285 most popular collections. Then, we propose heuristic algorithms to identify NFT wash trading/traders. At last, we conduct extensive experiments to measure the financial impact/trend/market liquidity of NFT wash trading, and present six findings from the marketplace design, profitability, NFT project design, payment token, user behavior, and NFT ecosystem.

In the future, we will develop a real-time visualization interface to track on-chain data, making it available for researchers. This tool aims to provide a more detailed and immediate understanding of NFT transactions and wash trading activities, further supporting research in this area. Additionally, we will collect more known instances of NFT wash trading behavior and design detection algorithms to cover a wider range of such activities. This will enhance our ability to identify and understand the full scope of wash trading in the NFT market.


\begin{acks}
This research/project is supported by the Open Research Fund of The State Key Laboratory of Blockchain and Data Security, Zhejiang University), the National Natural Science Foundation of China (62302534).
\end{acks}

    \balance
    \bibliographystyle{ACM-Reference-Format}
    \bibliography{ref}



\end{document}